\documentclass[twocolumn,usenames,dvipsnames]{svjour3}
\journalname{Applied Physics B}

\DeclareMathAlphabet{\mathpzc}{OT1}{pzc}{m}{it}

\usepackage{amsmath}
\usepackage{amssymb}
\usepackage{bm}
\usepackage{graphicx}
\usepackage[english]{babel}
\usepackage{tikz}
\newcommand\circledtext[1]{\protect\tikz[baseline=(char.base)]{
            \protect\node[shape=circle,draw,inner sep=1pt] (char) {#1};}}
\newcommand{\jmprevision}[1]{#1}

\usepackage{comment} 

 \usepackage[absolute]{textpos}
 \usepackage[author=Jordi]{pdfcomment}

\newcommand{\rvec}{\bm{r}}
\newcommand{\R}{\bm{R}}
\newcommand{\x}{\bm{x}}
\newcommand{\X}{\bm{X}}

\newcommand{\mumeter}[0]{\ensuremath{\mu\mathrm{m}}}
\newcommand{\ket}[1]{\ensuremath{|{#1}\rangle}}

\newcommand{\half}{\ensuremath{\frac{1}{2}}}

\newcommand{\mat}[1]{\ensuremath{\mathbb{#1}}} 

\newcommand{\talpha}{\ensuremath{\tilde\alpha}}
\newcommand{\fo}[1]{\ensuremath{f_{\mathrm{#1}}^{(0)}}}

\newcommand{\dip}{\ensuremath{\mathfrak{p}}}
\newcommand{\dipvec}{\ensuremath{\bm{\dip}}}

\newcommand{\om}{\omega}
\newcommand{\Om}{\Omega}
\newcommand{\com}{\textsc{com}}
\newcommand{\str}{\textsc{str}}
\newcommand{\omcom}{\om_{\com}}
\newcommand{\omcomz}{\om_{\com,0}}
\newcommand{\omstr}{\om_{\str}}
\newcommand{\omstrz}{\om_{\str,0}}
\newcommand{\omdip}{\om_{\mathrm{dip}}}
\newcommand{\mred}{\ensuremath{m_{\mathrm{red}}}}
\newcommand{\mtot}{\ensuremath{m_{\mathrm{tot}}}}

\begin{document}
 \title{Collective modes of a trapped ion-dipole system}
 \subtitle{Towards measuring, controlling and entangling electric dipoles with atomic ions}
 \author{Jordi Mur-Petit \and Juan Jos\'e Garc\'\i a-Ripoll}
 \institute{%
  J. Mur-Petit \at
   Instituto de Estructura de la Materia,
   IEM-CSIC, Serrano 123, E-28006 Madrid, Spain
   \\
   Kavli Institute for Theoretical Physics,
   University of California, Santa Barbara, CA 93106, USA
   \\
   \email{jordi.mur@csic.es}
  \and
  J. J. Garc\'\i a-Ripoll \at
   Instituto de F\'\i sica Fundamental, IFF-CSIC, Serrano 113 bis,
   E-28006 Madrid, Spain}
 \date{\today}


 \maketitle
 \begin{abstract}
 We study a simple model consisting of an atomic ion and a polar molecule trapped in a single setup, taking into consideration their electrostatic interaction.
 We determine analytically their collective modes of excitation as a function of their masses, trapping frequencies, distance, and the molecule's electric dipole moment.
 We then discuss the application of these collective excitations to cool molecules, to entangle molecules and ions, and to realize two-qubit gates between them. We finally present a numerical analysis of the possibility of applying these tools to study magnetically ordered phases of two-dimensional arrays of polar molecules, a setup proposed to quantum-simulate some strongly-correlated models of condensed matter.
 

 \end{abstract}


\section{Introduction}
\label{sec:intro}

\begin{quote}
\textit{%
Experimental physics is the art of observing the structure of matter and of
detecting the dynamic processes within it. But in order to understand the
extremely complicated behaviour of natural processes as an interplay of a
few constituents governed by as few as possible fundamental forces and
laws, one has to measure the properties of the relevant constituents and
their interaction as precisely as possible. And as all processes in nature are
interwoven one must separate and study them individually.%
}%
\hfill
Wolfgang Paul
\end{quote}

\noindent %
This citation from W. Paul's Nobel lecture in 1989~\cite{paul1989nobel} expresses in a clear manner the importance of measurement in physics, and the advantages that come when measurement can be done on single particles and for extended periods of time.
Actually, these ideas can be regarded as the \textit{motto} behind the work of Paul, Dehmelt, and Ramsey, who were awarded the Nobel Prize in Physics that year, ``for the invention of the separated oscillatory fields method'' the latter, and the ion trap the former two~\cite{paul1989nobel,dehmelt1989nobel,ramsey1989nobel}.

In recent years, the achievement of these goals has seen spectacular progress thanks to a steady improvement of the trapping methods for individual particles, both charged and neutral.
In this context, the use of ions confined in Paul traps has proved particularly 
valuable, leading for example to extremely precise measurements of time and frequency~\cite{ludlow2008,rosenband2008} or extremely weak forces~\cite{biercuk2010,murpetit2012}, as well as to proposals to measure the electric dipole moment (EDM) of the electron with heavy molecular ions~\cite{leanhardt2011,cossel2012}.
In the near future, one expects even more precise measurements, approaching the Heisenberg limit, to be accomplished thanks to the exquisite degree of control at the single-quantum-state level of these individual particles~(see e.g.~\cite{blatt2012,bloch2012nphys}), together with the development of quantum metrology, i.e., measurement strategies that take advantage of the quantum nature of the probes used~\cite{banaszek2009,giovannetti2011}.

The degree of accuracy with which single trapped ions can be measured and controlled has lead them to become one of the most advances technologies toward the realization of Quantum Information Processing~\cite{haeffner2008} and has already given some remarkable results in Quantum Simulation~\cite{blatt2012}. In this context, a proper understanding and control of the \textit{collective} motion of trapped ions due to their electrostatic interactions has been a key ingredient, enabling the realization of cooling of co-trapped ions~\cite{kielpinski2000,vogelius2006}, two-qubit gates~\cite{leibfried2003,schmidt-kaler2003} and the creation of two- and many-particle entanglement~\cite{turchette1998,sackett2000}.

Beyond charged particles, cold polar molecules have also attracted a lot of interest because of their foreseen applicability to 
measure variations in the values of fundamental constants~\cite{flambaum07}, parity- and time-violating interactions~\cite{demille2008}, to implement
quantum information~\cite{demille2002,andre2006,pellegrini2011}, as well as to study and control  chemical reactions at ultralow temperatures~\cite{krems2008,bell09b}.
Polar molecules are also elementary units of recent proposals for quantum simulators of strongly-correlated condensed-matter systems, including quantum magnetism models~\cite{andre2006,buechler2007,gorshkov2011} (see also~\cite{lahaye2009}).
Establishing a bridge between the fields of trapped ions and neutral particles, 
in the last years a growing number of groups have reported the creation of hybrid systems with ions and neutral atoms, which have enabled the study of ultra-cold collisions~\cite{grier2009,zipkes2010,schmid2010} and cooling~\cite{ravi2012,sivarajah2012,haze2013}, ultra-low-energy chemical reactions~\cite{rellergert2011,hall2011,deiglmayr2012}, etc. We think that hybrid systems mixing atomic ions and cold mol\-e\-cules will be the next step along this road.

Building on the experience with atomic and molecular ions, where the determination of normal modes of motion~\cite{kielpinski2000,james1998} has proved so important, we address here the determination of the collective modes in a system consistent of an ion and an electric dipole, independently confined in harmonic traps. In Sect.~\ref{sec:eigenmodes} we present the basic setup under consideration, find the new equilibrium positions of the two particles when their electrostatic interaction is taken into account, and determine analytically the eigenfrequencies of their collective motion. We also perform an analysis (Sect.~\ref{ssec:stability}) of the stability of the system against collapse of the ion onto the dipole, or its ejection from the trap. Then, in Sect.~\ref{sec:appl} we present a number of immediate applications of these results: molecular cooling, measurement and control of EDMs (see also~\cite{iondipoleshort}), atom-molecule entanglement, and mapping ordered phases of dipoles on a lattice via a technique we name ``ion-dipole force microscopy'' (IDFM). Finally, we draw our conclusions and provide some outlook of this work in Sect.~\ref{sec:outlook}.


\section{Eigenmode description of a coupled ion-dipole system}
\label{sec:eigenmodes}

\subsection{A hybrid ion-dipole system}
\label{ssec:system}

\begin{figure}[tb]
  \centering
  \includegraphics[width=0.8\columnwidth]{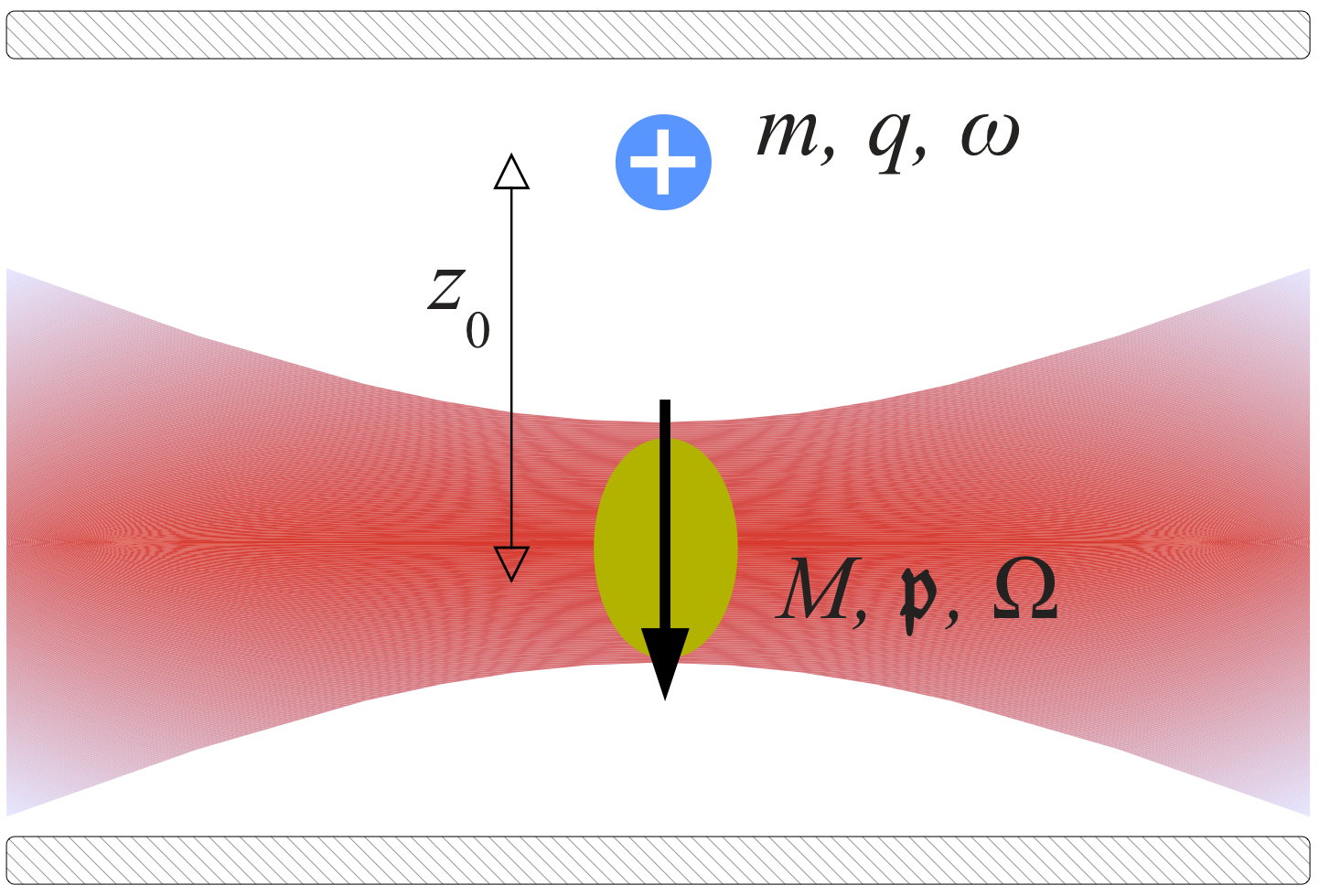}
  \caption{{\em Scheme of the system.} An ion of mass $m$ and charge $q$ (blue circle with white ``+'') is kept in a Paul trap (horizontal hatched bars) while a polar molecule (yellow oval) of charge $M$ and EDM \dipvec\ (thick arrow) is confined by a focused laser beam (red shaded area) a distance $z_0$ beneath. $\om$ and $\Om$ stand for the corresponding trapping frequencies.}
  \label{fig:system}
\end{figure}

We start considering a system composed of an ion of mass $m$ and charge $q$ confined in a harmonic trap of trapping frequency $\om$, and an electric dipole \dipvec\ of mass $M$ in a harmonic trap of trapping frequency $\Om$, as depicted in Fig.~\ref{fig:system}.
For simplicity, we assume that each trap has spherical symmetry around its minimum, that we take as origin of coordinates for the dipole, and at position $\x_0$ for the ion.
The ion-dipole interaction couples the two particles, so that the total energy is written as
\begin{equation}
  W
  = \half m \om^2 (\x-\x_0)^2
  + \half M \Om^2 \X^2
  + \frac{q\dipvec\cdot
     \left( \x-\X \right)}{4\pi\epsilon_0|\x-\X|^3}
  \label{eq:energy}
\end{equation}
Here, $\bm{x}$ is the position vector of the ion, $\bm{X}$ the position of the dipole.
We can simplify the expression for $W$ by going to the coordinate system defined by the relative, $\rvec=\x-\X$, and center of mass (c.o.m.), $\R=(m\x + M\X)/(m+M)$, coordinates:
\begin{eqnarray}
  W
  &=&
   \half m \om^2 (\R-\bm{x}_0)^2
  + \half M \Om^2 \R^2
  \nonumber \\
  &+& \half \mred \omstrz^2 \rvec^2
  - \mred \om^2 \x_0\cdot\rvec
  + \frac{q\dipvec\cdot\rvec}{4\pi\epsilon_0|\rvec|^3}
  \nonumber \\
  &+& \mred(\om^2-\Om^2) \R\cdot\rvec \:,
 \label{eq:energyrR}
\end{eqnarray}
where we introduced
 $\mtot=m+M$ as the total mass
 and the reduced mass $\mred=m M/\mtot$,
 and we identified the relative-motion collective mode (``stretch mode'') frequency for the uncoupled ($\dip=0$) and overlapping ($\x_0=0$) system: $\omstrz^2 := (m\Om^2 + M\om^2)/\mtot$. Below, it will become useful that we also define the ``center-of-mass mode'' frequency of the same system, $\omcomz^2 := (m\om^2 + M\Om^2)/\mtot$.
\begin{figure}[tb]
  \centering
  \begin{minipage}{0.6\columnwidth}
    \includegraphics[width=\columnwidth]{
       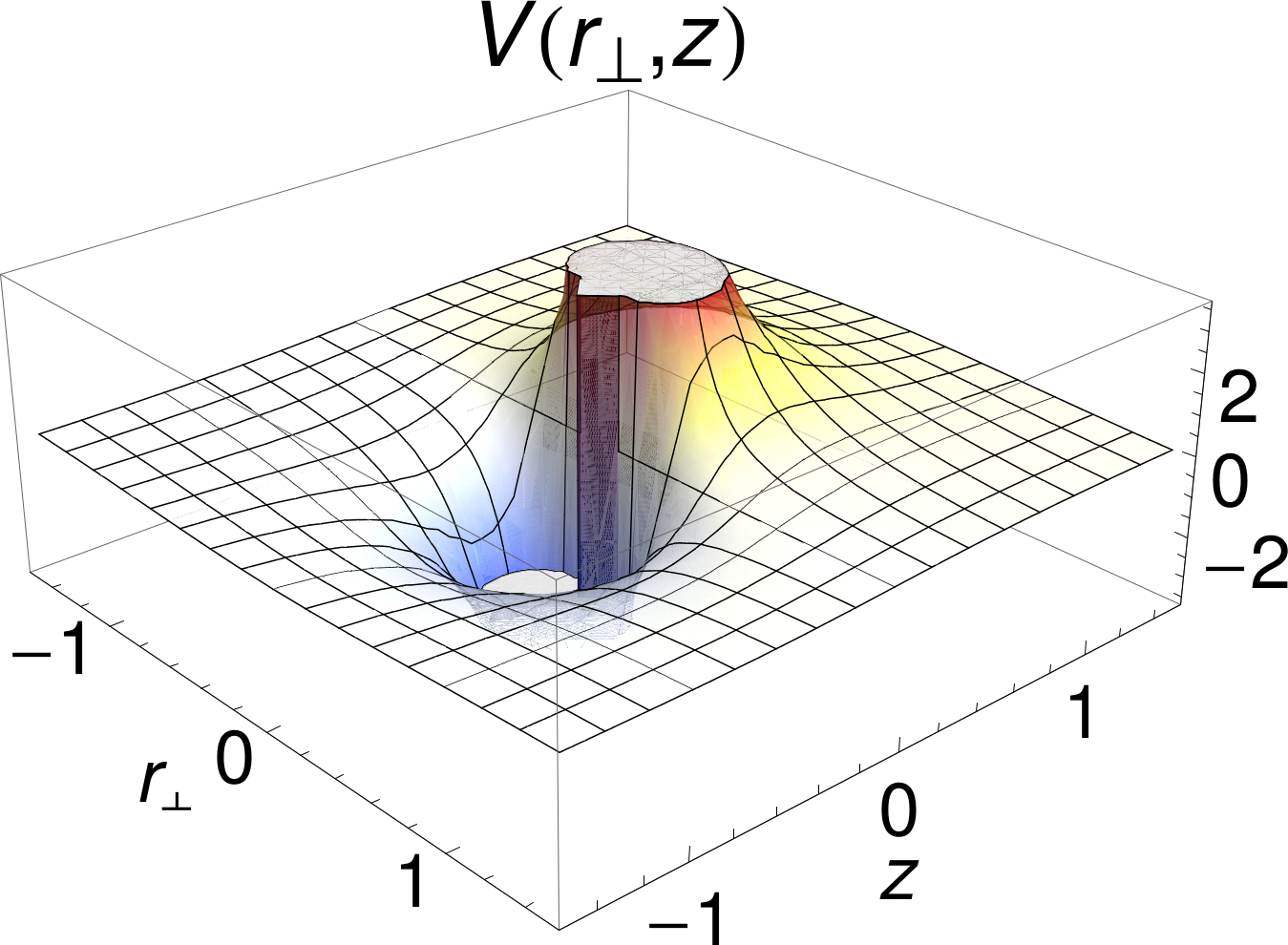}
  \end{minipage}%
  \begin{minipage}{0.35\columnwidth}
    \includegraphics[width=\columnwidth]{
       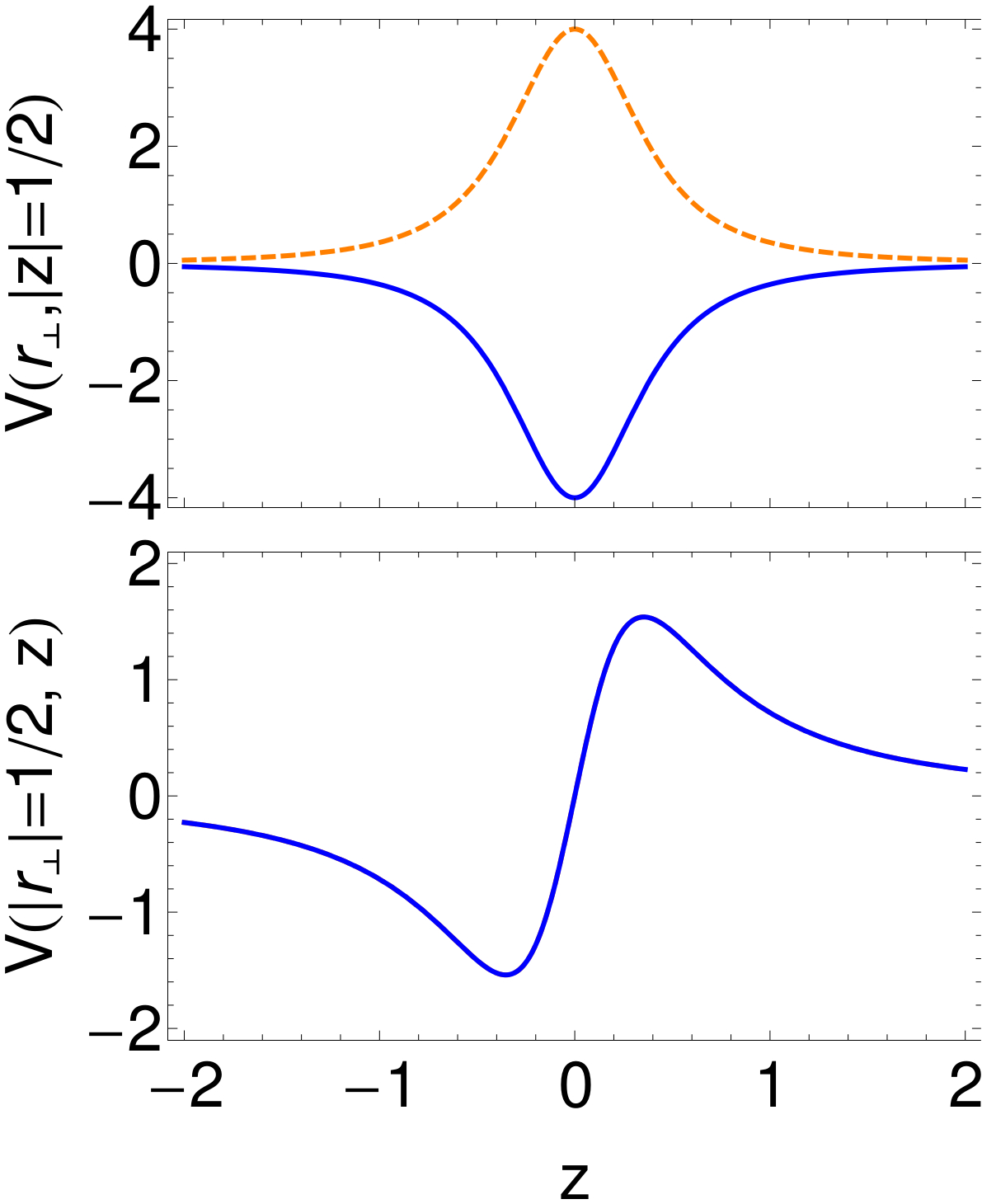}
  \end{minipage}
  \caption{%
    (Left) Ion-dipole electrostatic energy, $V=\bm{r}/r^{3}$, $r_{\perp}=\sqrt{x^2+y^2}$.
    (Right, top) Cuts of $V$ along $z=+1/2$ (dashed orange line)
      and $z=-1/2$ (solid blue).
    (Right, bottom) Cuts of $V$ along $r_{\perp}=\pm1/2$ (both lines overlap).%
    }
  \label{fig:coupling}
\end{figure}

The first two terms in Eq.~(\ref{eq:energyrR}) indicate the center of mass coordinate to be in a double harmonic potential, with one minimum at $\x_0$ and the other at the origin, while the terms on the second line correspond to the relative coordinate being in a harmonic potential displaced from the origin of coordinates, while the coupling with \dip\ amounts to a potential further displacing it. The last term indicates that there is coupling between relative and c.o.m.\ if {\em and only if} the two trapping frequencies differ.

In the following paragraphs, we will find out the new equilibrium values for $\rvec$ and $\R$ taking into account these couplings and the effect of the ion-dipole interaction [last term on the second line of Eq.~(\ref{eq:energyrR})], and determine the normal modes (collective excitation) frequencies corresponding to small oscillations about these new minima.


\subsection{New equilibrium positions}
\label{ssec:neweq}
The new equilibrium configuration will be determined by the values of $\rvec$, $\R$ that minimize $W$ in Eq.~(\ref{eq:energyrR}).
One can get a certain intuition on the behaviour of this function by analyzing the coupling term, $\propto \rvec/|\rvec|^3$. This is drawn in Fig.~\ref{fig:coupling}. Looking at the lower inset, which shows $W$ for fixed $r_\perp=\sqrt{x^2+y^2}$, we see that the energy is minimized for a finite $z$ (negative, meaning in the direction opposite to the dipole's arrow).
On the other hand, for fixed $z$ (top inset), the equilibrium solution is $x=y=0$, with the point being unstable if $z>0$ and stable if $z<0$. This is understood taking into account that the ``head'' of the dipole corresponds to the positive charge and, hence, to the side of the dipole that most repels the ion.

\begingroup
\begin{table}[b]
  \caption{Typical values for trapped ion and dipole systems, and derived parameters $L$,$\alpha$ for candidate ion-dipole systems at the reported distances $z_0$.
  Ground state EDMs (in Debye) and linewidths (in $2\pi$ MHz) taken from indicated references. 
    }
  \label{tab:frequencies}
  \begin{tabular}{lllll} 
  \hline\hline
  Ion             & $q$ & $\Gamma_{\mathrm{ion}}$   & $\om/2\pi$ \\
  ~$^{40}$Ca$^+$ & +e  &  20 \cite{james1998}  & 1 MHz & \\
  \hline
  Dipole              & $\dip$ (D)             & $\Gamma_{\mathrm{dip}}$
    & $\Om/2\pi$ & $L$ (\mumeter)
  \\
  ~$^{40}$K$^{87}$Rb & 0.566 \cite{ni2008sci}  & 0.50 \cite{aikawa2009}
    & 1 kHz & 4.3
  \\
  ~$^{40}$Ca$^{1}$H  & 2.94 \cite{steimle2004} & 2.74 \cite{berg1996}
    & 1 kHz & 8.5
  \\  \hline
  Hybrid syst. &
    $z_0$ ($\mumeter$) & $\alpha/2\pi$ & $\alpha/\sqrt{\om\Om}$
      & $\talpha$ \\
  ~Ca$^+$-KRb
    & ~1 & 4.10 MHz & 130   & $1.0(-3)$ \\
    & 10 & 41.0 kHz & 1.30  & $1.0(-7)$ \\
    & 20 & 10.3 kHz & 0.324 & $6.5(-9)$ \\
  ~Ca$^+$-CaH
    & ~1 & 21.3 MHz & 674  & $5.4(-3)$ \\
    & 10 &  213 kHz & 6.74 & $5.4(-7)$ \\
    & 20 & 53.3 kHz & 1.68 & $3.4(-8)$ \\
  \hline\hline
\end{tabular}
\end{table}
\endgroup

Determining the new equilibrium positions, 
$\rvec_*=(x_*,y_*,z_*)$, $\R_*=(X_*,Y_*,Z_*)$,
is in general a rather complex problem, 
and we will make a few approximations in order to obtain analytic solutions.
For simplicity, we start taking the position of the ion to define our $z$ axis, i.e., $\x_0=(0, 0, z_0)$, cf.\ Fig.~\ref{fig:system}.
Further, we assume that the dipole is oriented along the ion-dipole axis, $\dipvec = (0, 0, \dip)$.
The minimum of the energy is then given by one of the following solutions: \begin{subequations}
\begin{eqnarray}
\left\{ %
 \begin{array}{rl}
   x_*=y_* &=  L \sqrt{ \left(1-\frac{L^2}{9z_0^2} \right) /2 }, \\
   z_*     &=  L^2/(3z_0) , \\
   X_*=Y_* &= -x_* (\om^2-\Om^2)/\omcomz^2 , \\
   Z_* \omcomz^2  &= z_0 \om^2 m/\mred -z_*(\om^2-\Om^2)
 \end{array} \right\}  \:;
 \label{eq:neweq1}
\\      
\left\{ %
 \begin{array}{rl}
   x_*=y_* &=  0 , \\
   z_*: &  \mathrm{~solution~of:~} z^3(z-z_0)=2L^4 , \\
   X_*=Y_* &=  0 , \\
   Z_* \omcomz^2  &= z_0 \om^2 m/\mred -z_*(\om^2-\Om^2)
 \end{array} \right\}  \:.
 \label{eq:neweq2}
\end{eqnarray}
\label{eq:neweq}%
\end{subequations}
Here we have defined a characteristic length of the 
interaction by
$L^4 = q\dip/(4\pi\epsilon_0)(m\om^2 + M\Om^2)/(mM\om^2\Om^2)$; for typical values of ultracold, trapped systems (cf. Table~\ref{tab:frequencies}) we have $L \sim 1-10~\mumeter$.

\begin{figure}[bt]
  \centering
  \includegraphics[width=0.8\columnwidth]{
     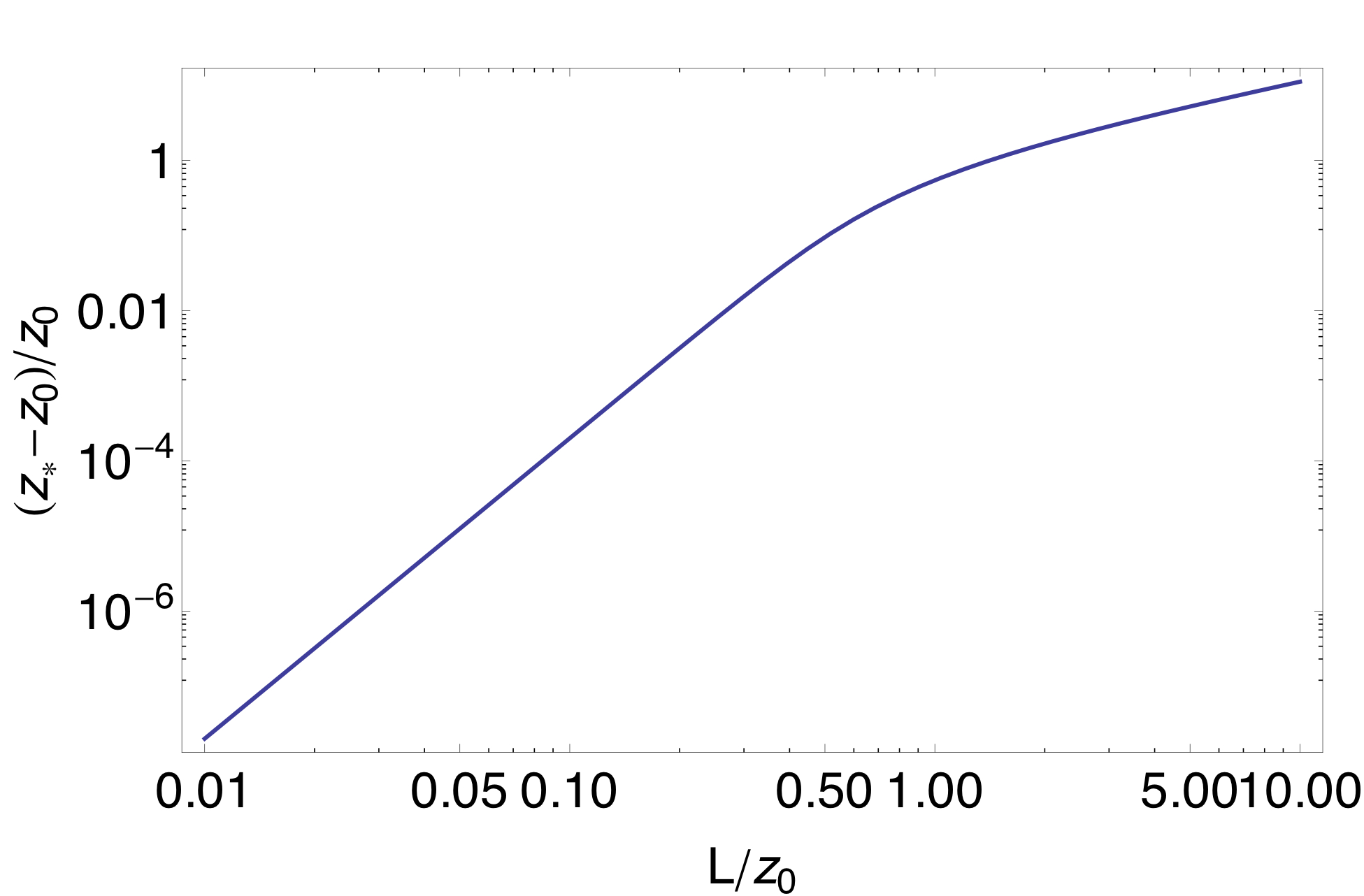}
  \caption{New equilibrium position as a function of coupling strength, given in terms of $L$, as predicted by Eq.~(\ref{eq:neweq2}).}
  \label{fig:displacement}
\end{figure}

From the solutions (\ref{eq:neweq1}) and (\ref{eq:neweq2}), only the second one fulfills the requirement that $z_* \to z_0$ when we approach the uncoupled system ($\dip \to 0$ or $L \to 0$). For this solution, we show in Fig.~\ref{fig:displacement} the displacement of the relative coordinate equilibrium position, $z_*-z_0$, as a function of the coupling length: note how it is very small for $L\lesssim z_0/2$, and then grows linearly,
$z_*-z_0 \propto L \propto \dip^{1/4}$.
%


\subsection{Collective eigenmodes of the system}
\label{ssec:eigen}

In the previous subsection, we have determined the new equilibrium positions of the ion-dipole system due to the electrostatic interaction between them. Now, we will find the collective excitation frequencies of the system. To this end, we follow~\cite{kielpinski2000} and expand the energy $W(\rvec,\R)$ in a Taylor series around $\rvec_*$ and $\R_*$ given by Eq.~(\ref{eq:neweq2}).
Keeping only terms up to quadratic order in the displacements, we can write the expansion in matrix form as
\begin{eqnarray}
  W = \frac{1}{2}m\om^2 \, \mat{R} \cdot \mat{A} \cdot \mat{R}^\mathfrak{t}
\end{eqnarray}
where $\mat{R}$ is the 6-component vector of displacements, $\mat{R}=(\rvec-\rvec_* ,~ \R-\R_*)^\mathfrak{t}$, $\mathfrak{t}$ means transpose, and $\mat{A}$ is the dimensionless matrix
\begin{eqnarray}
 \mat{A}
 &=
 \left( %
 \begin{array}{cccccc}
   A & 0 & 0 & D & 0 & 0 \\
   0 & A & 0 & 0 & D & 0 \\
   0 & 0 & B & 0 & 0 & D \\
   D & 0 & 0 & C & 0 & 0 \\
   0 & D & 0 & 0 & C & 0 \\
   0 & 0 & D & 0 & 0 & C
 \end{array} \right)
\end{eqnarray}
where $A = (\mred/m)(\omstrz/\om)^2$,
$B = A -30\talpha$,
 $C = (\mtot/m)(\omcomz/\om)^2$,
and $D = (\mred/m)[1-(\Om/\om)^2]$. 
Here, we defined
$\alpha = q\dip/(4\pi\epsilon_0z_0^2)$,
which has units of energy,
and
$\talpha = \alpha/(m \omega^2z_0^2)$
which is dimensionless and gives an estimate of the ion-dipole coupling energy vs.\ trapping energy at the interparticle distance $z_0$.
For typical values (Table~\ref{tab:frequencies})
we find  $\alpha/\sqrt{\hbar^2\om \Om} \sim 0.1-100$ (see Sect.~\ref{ssec:idfm} below), which means that one can reach a regime of strong coupling due to the ion-dipole interaction.
For later convenience, we introduce also a characteristic frequency related to the interaction by
\begin{equation}
  \omdip^2 := 30\frac{\alpha}{\mred z_0^2}
  \label{eq:omdip}
\end{equation}
so that $B\equiv(\mred/m)(\omstrz^2-\omdip^2)/\om^2$. 

From the structure of $\mat{A}$ we see immediately two things:
(i) There is a decoupling of the three spatial directions, i.e., $x$ is only coupled with $X$, $y$ with $Y$, and $z$ with $Z$.
(ii) When $\Om=\om$ ($D=0$), the eigenfrequencies are given directly by $\omstrz$ and $\omcomz$ for coordinates $x,y$ (with mass $\mred$), and $X,Y,Z$ (with $\mtot$), respectively. For $z$, it depends on $\omstrz^2-\omdip^2$; we discuss this in more detail in Sect.~\ref{ssec:stability} below.

For the general case (any $\Om/\om$), it is easy to find the eigenfrequencies of the system. To this end, we decompose $\mat{A}$ in the form
$\mat{A} = \mat{K}^T \mat{D} \mat{K}$, where $\mat{D}$ is a diagonal matrix, whose elements directly give the excitation modes of the system, and the columns of $\mat{K}$ provide the corresponding eigenvectors.
The final result for the eigenvalues related to the $(z,Z)$ coordinates (the only ones involved with the dipole moment) is given by
\begin{eqnarray}
  \mat{D}_{\pm} &=& \tilde{r} \pm \sqrt{\tilde{s}} \,, 
  \label{eq:eigenvalues}
\end{eqnarray}%
where we defined
\begin{subequations}
\begin{eqnarray}
  \tilde{r} &=&
     \frac{1 +2\mu +2\mu^2 + \mu\nu^2(2+2\mu+\mu^2)}{2(1+\mu)^2}
     +15 \talpha \:, \\
  \tilde{s} &=& \tilde{r}^2
     -\left[ \mu\nu^2 + 30\talpha(1+\mu\nu^2) \right] \:.
\end{eqnarray}
\label{eq:rs}%
\end{subequations}
Here, 
$\mu:=M/m$, and $\nu:=\Om/\om$ are dimensionless parameters that characterize the system. A study of the eigenvectors for the case $D\to 0$ (i.e., $\Om\to\om$) allows us to see that the \com\ mode is related to $\mat{D}_+$ 
and the \str\ mode to $\mat{D}_-$. 
Hence, the frequencies of the normal modes can be expressed in general form via 
\begin{subequations}
\begin{eqnarray}
  \omcom
  &=& \om \sqrt{\frac{m}{\mtot} \mat{D}_+} 
   =  \om\sqrt{\frac{\tilde{r}+\sqrt{\tilde{s}}}{1+\mu}} 
  \:, \\
  \omstr
  &=& \om \sqrt{\frac{m}{\mred} \mat{D}_-} 
   =  \om\sqrt{(\tilde{r}-\sqrt{\tilde{s}})\frac{1+\mu}{\mu}} 
  \:.
\end{eqnarray}
\label{eq:eigenfreqs}%
\end{subequations}
Eqs.~(\ref{eq:eigenvalues}-\ref{eq:eigenfreqs}) 
contain as limits $\omcomz$ and $\omstrz$ for the uncoupled and overlapping system [as defined after Eq.~(\ref{eq:energyrR})], and allow to calculate the normal mode frequencies for the fully general case
$m\neq M$, $\om\neq\Om$, $\x_0\neq0$.
Their particular values when ion and dipole have the same mass, $\mu=1$, and trapping frequency, $\nu=1$, read
\begin{subequations}
\begin{eqnarray}
 \omcom/\om
 &=&
 \sqrt{ 5+60\talpha + 3|1-20\talpha| }/2\sqrt{2},
 \nonumber \\
 &\longrightarrow&
 \left\{ \begin{array}{llc}
   1                \,, & ~~~ \talpha \to 0      & ~~* \\
   \sqrt{15\talpha} \,, & ~~~ \talpha \to \infty & ~~\dagger
 \end{array} \right.
 \\
 \omstr/\om
 &=&
 \sqrt{ 5+60\talpha - 3|1-20\talpha| }/\sqrt{2},
 \nonumber\\
 &\longrightarrow&
 \left\{ \begin{array}{llc}
  \sqrt{1+60\talpha} \,, & ~~~ \talpha \to 0      & ~~*\\
   2 \,,  & ~~~ \talpha \to \infty & ~~\ddagger
 \end{array} \right.
\end{eqnarray}
\label{eq:limits}%
\end{subequations}
Thus, in this particular case ($\mu=\nu=1$), and for vanishing coupling, \com\ and \str\ become degenerate and equal $\om$.
Physically this just means that in abscence of interaction the motion of the ion does not affect the dipole and viceversa.
We note that, numerically, the limit $\Om\to\om$ has to be taken with care as the two normal modes involve both of $z$ and $Z$ for all $\nu\neq1$, while exactly at $\Om=\om$ these coordinates decouple, as discussed earlier.
This is most apparent when studying the eigenvectors, which can be written in the (unnormalized) form $\psi_{\mathrm{com,str}}=(u_{\mathrm{com,str}}, 1)$ for any $\nu\neq1$, but they read $\psi_{\mathrm{str}}\equiv(1,0)$, $\psi_{\mathrm{com}}\equiv(0,1)$ for $\nu=1$. In fact, $u_{\mathrm{str}} \propto 1/(\nu-1) \to \infty$, $u_{\mathrm{com}}\propto(\nu-1)\to0$: the pathological character of the limit is clear in $u_{\mathrm{str}}$.

\begin{figure}[bht]
  \centering
  \textbf{(a)}
  \raisebox{-0.8\height}{%
  \includegraphics[width=0.77\columnwidth]{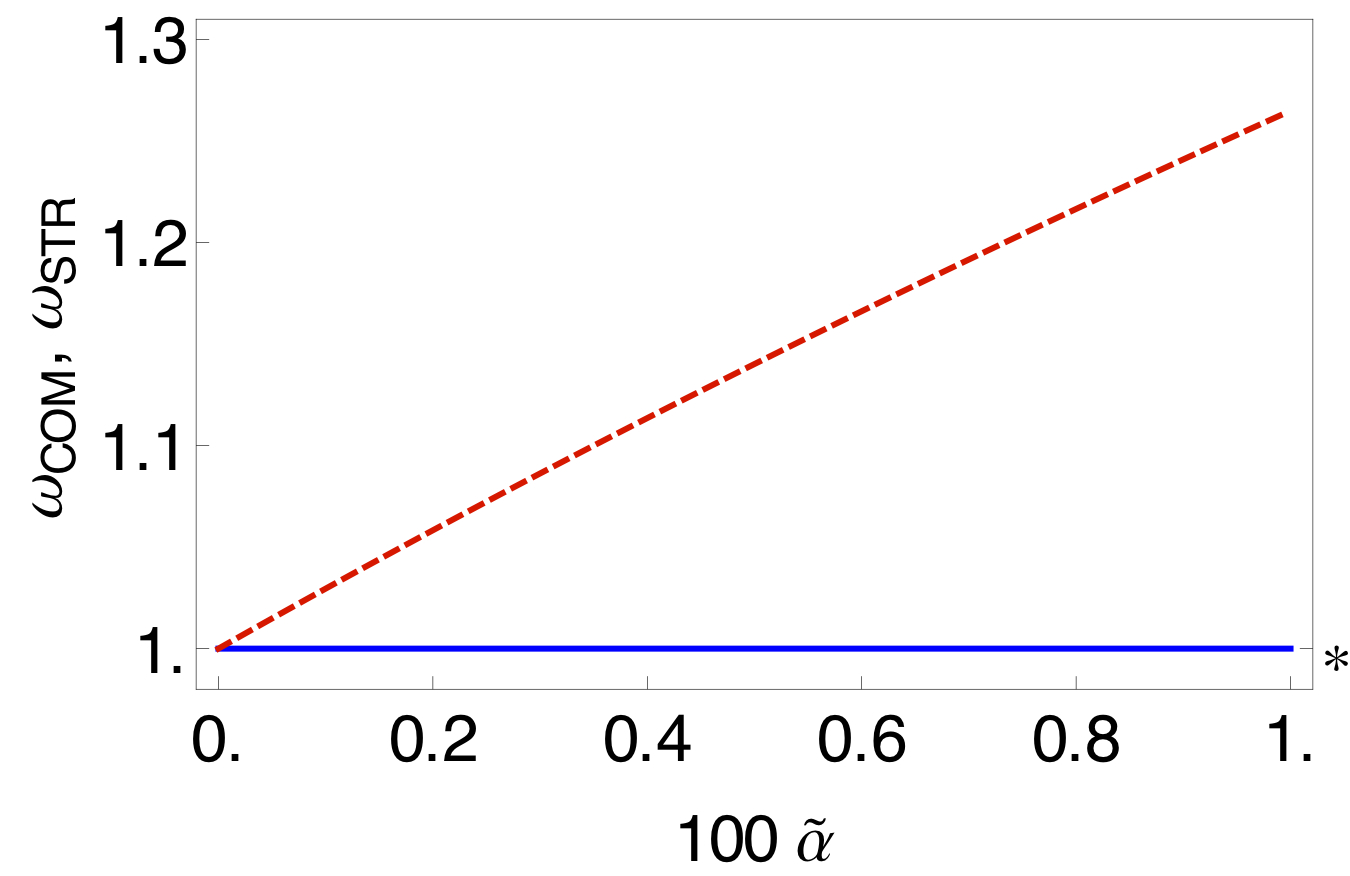}}
  
  \textbf{(b)}
  \raisebox{-0.8\height}{%
  \includegraphics[width=0.77\columnwidth]{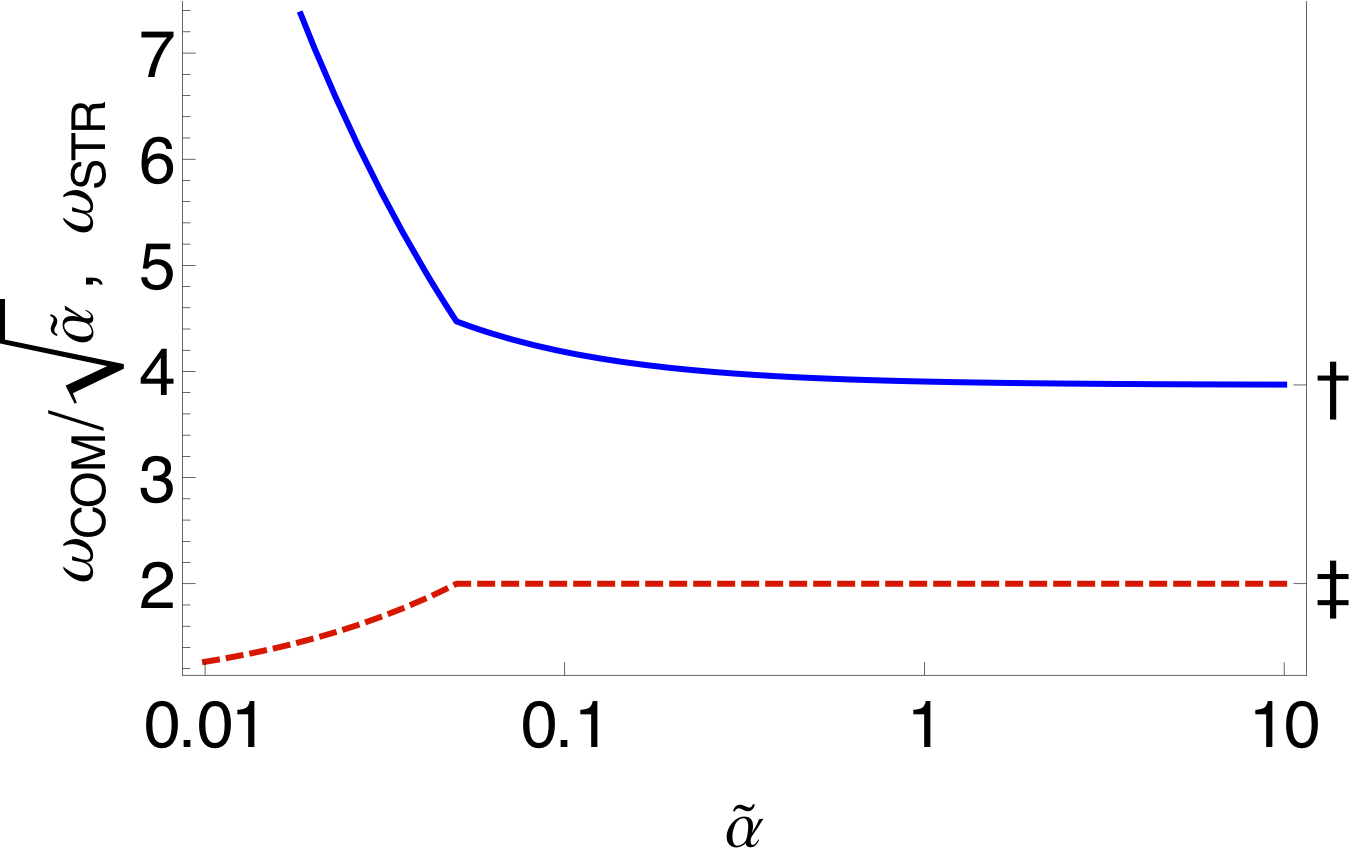}}

  \textbf{(c)}
  \raisebox{-0.8\height}{%
  \includegraphics[width=0.77\columnwidth]{
     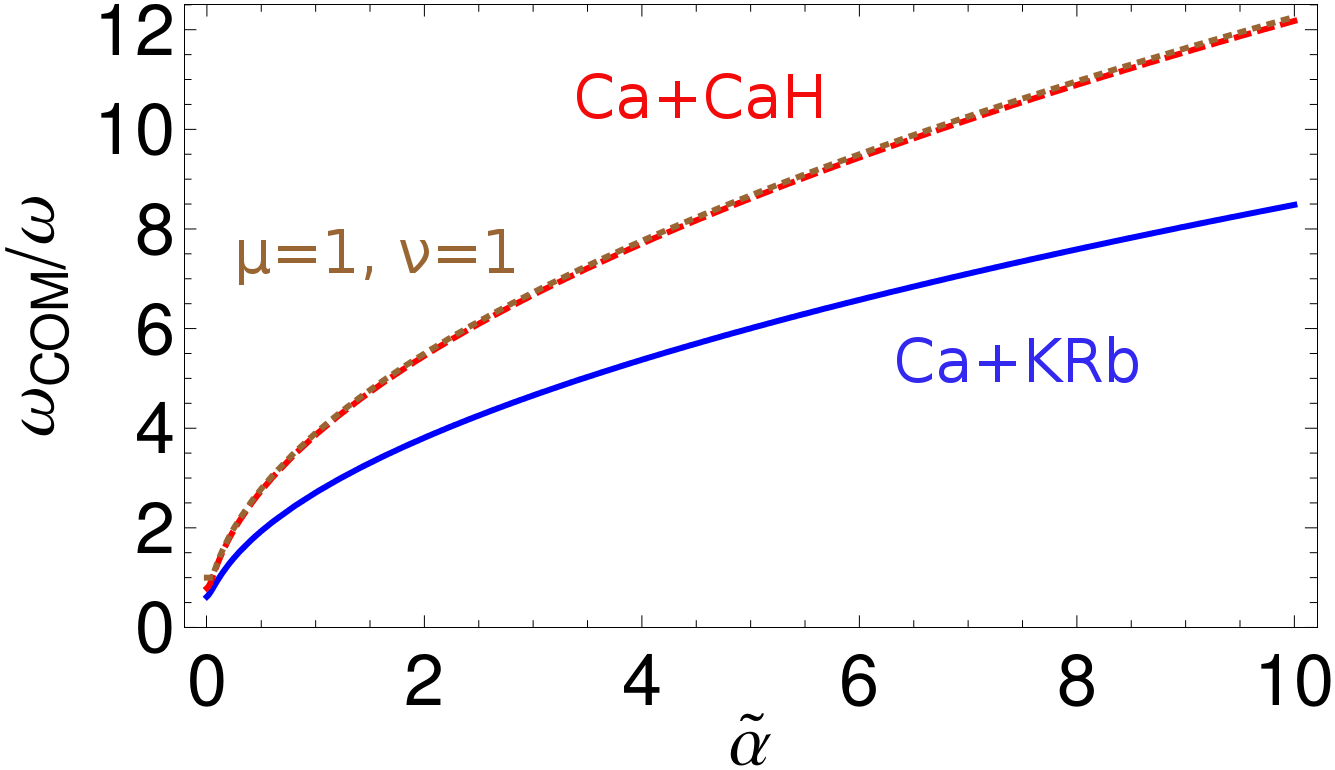}}

  \textbf{(d)}
  \raisebox{-0.8\height}{%
  \includegraphics[width=0.77\columnwidth]{
     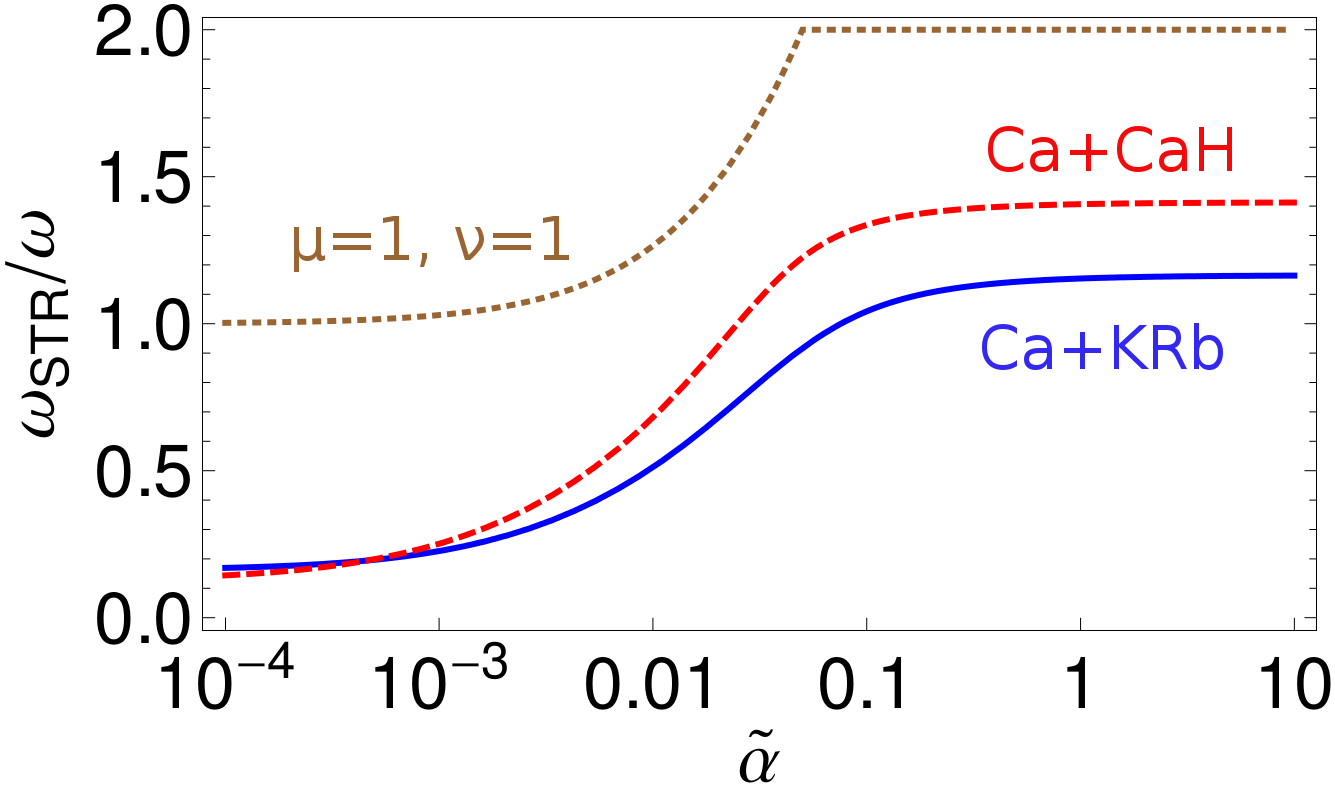}}
  \caption{
    \jmprevision{%
    Eigenfrequencies $\omcom/\om$ (blue, solid lines) and $\omstr/\om$ (dashed red) for $\mu=\nu=1$:
    \textbf{(a)} for small $\talpha\to0$;
    \textbf{(b)} limiting behaviours for $\talpha\to\infty$.
    Symbols on the right side $(*,\dagger,\ddagger)$ signal the analytic limits in Eqs.~(\ref{eq:limits}).
    \textbf{(c)} $\omcom/\om$ and \textbf{(d)} $\omstr/\om$ as a function of $\talpha$ for $\nu=0.1$ and the mass ratio corresponding to Ca$^+$+KRb (blue solid) and Ca$^+$+CaH (red dashed); for comparison we include also the curves for $\mu=1,\nu=1$ (brown dotted).
    Note the different scales for $\talpha$ on the different panels.
    }
  }
  \label{fig:modes}
\end{figure}

At strong coupling, \com\ becomes increasingly stiffer, while the \str\ mode frequency saturates. The behaviour of these functions is shown in Fig.~\ref{fig:modes} for a range of parameters. In Fig.~\ref{fig:modes}(a,b), we present $\omcom$ and $\omstr$ for the case of equal masses and trapping frequencies, $\mu=1,\nu=1$, when the coupling is small [Fig.~\ref{fig:modes}(a)] and large [Fig.~\ref{fig:modes}(b)]. We recover the degeneracy of the two modes at vanishing coupling (indicated by the star \raisebox{-0.3\height}{*}  on the right side of the plot), while $\omstr$ separates linearly from this limit for small $\talpha$, as predicted by Eqs.~(\ref{eq:limits}). On the other hand, $\omcom/\sqrt{\talpha}$ and $\omstr$ converge quickly to their strong-coupling limits (shown by $\dagger$ and $\ddagger$).
The dependence of $\omcom$ and $\omstr$ on coupling strength for more realistic values is shown in Figs.~\ref{fig:modes}(c) and (d), respectively, taking $\nu=1/10$ and two mass ratios, corresponding to systems composed of a Ca$^+$ ion and either KRb (solid lines) or CaH (dashed) molecules, which in their absolute ground states have EDMs  $\dip_{\mathrm{KRb}}=0.566$~D~\cite{ni2008sci} and $\dip_{\mathrm{CaH}}=2.94$~D~\cite{steimle2004}. 


\subsection{Stability}
\label{ssec:stability}

From the form of $\mat{A}$, we observe that even for the case of equal ion and dipole trapping frequencies, $\nu=1$, the electrostatic interaction between them can lead to the system being unstable to small fluctuations from the equilibrium configuration. This will depend on the relative magnitude of $\omdip$ and $\omstrz$. In the general case, we can summarize the different possibilities as follows:
\begin{itemize}
\item[$\bullet$] $q\dip<0$: this condition is equivalent to $\alpha<0$ and $\omdip^2<0$, and corresponds to a positive ion being in the direction opposite to the dipole's arrow (as in Fig.~\ref{fig:system}, cf. Sect.~\ref{ssec:neweq}), or a negative ion exactly ``on top'' of the dipole. In this case $B>0$ and small displacements about  $(z_*,Z_*)$ result in stable, harmonic oscillations.
\item[$\bullet$] $q\dip>0$ (i.e., $\omdip^2>0$). There are two possibilities:
  \begin{itemize}
  \item[$\star$] $\omstrz^2>\omdip^2$: again, $B>0$ and displacements lead to stable oscillations, though with a reduced oscillation frequency, $\sqrt{\omstrz^2-\omdip^2}$. This stabilization against the electrostatic repulsion, which allows to put a positive ion ``on top'' of the dipole arrow, is due to the ``trapping'' of the relative coordinate ($\omstrz\neq0$).
  \item[$\star$] $\omstrz^2<\omdip^2$: $B<0$ and equilibrium is unstable, with displacements in $z$ growing exponentially. In this case, the trapping of $\rvec$ is insufficient to counteract the electrostatic repulsion.
  \end{itemize}
\end{itemize}


\section{Applications}
\label{sec:appl}

The capacity to control the coupling between the internal (e.g., electronic, fine or hyperfine state) and external (translational or trapping) motion of {\em two ions} was proposed long time ago as a tool to cool the former~\cite{wineland1979}. This was experimentally realized between two atomic ions back in 1995~\cite{monroe1995}, and more recently the internal rotational state of molecular ions has also been cooled by controlled interactions with atomic ions~\cite{vogelius2006}, building on the knowledge of the collective modes of ion chains with different masses~\cite{james1998,kielpinski2000}.
This control has also enabled the realization of quantum gates between atomic ions~\cite{leibfried2003,schmidt-kaler2003}, and we proposed recently a similar scheme to control and entangle molecular and atomic ions~\cite{murpetit2012}.
In the next paragraphs, we discuss briefly some applications that result from  the knowledge of the collective modes of a system composed of an ion and an electric dipole.
As a first relevant example, we mention the possibility to cool the internal degrees of freedom (d.o.f.) of a polar molecule that are related to its EDM, such as its rotational state.

\subsection{Molecule cooling}
\label{ssec:cooling}

\begin{figure}[tb]
  \centering
  \includegraphics[width=0.9\columnwidth]{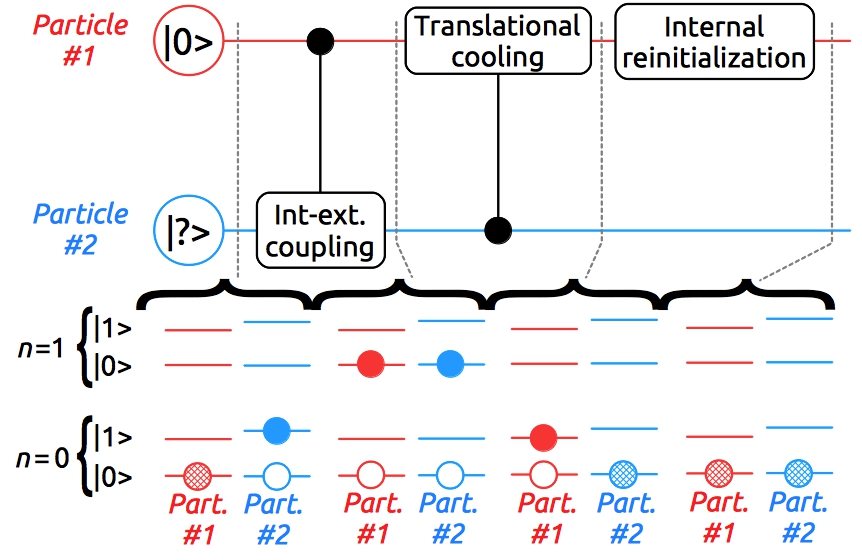}
  \caption{\textit{Cooling protocol via collective modes.}
  The top panel shows the protocol as a diagram in a quantum circuit with the steps discussed in the text indicated as gates.
  The bottom panel shows the energy states of the system at the start of the process and after each step (dashed vertical lines).
  The state of a collective mode is indicated by $n=0,1$, while the internal state of each particle is \ket{0} or \ket{1} as shown.
  Filled (empty) circles represent the state of each particle when particle \#2 is initially internally excited (in its ground state); the hatched circles are states common to both cases: regardless of \#2's initial state, the final state always has both \#1 and \#2 in their internal ground states.
  Note that the first two steps, even though we act only on one of the particles, involve the collective modes, which we indicate in the top panel by the vertical lines with a bullet.}
  \label{fig:cooling}
\end{figure}

The key ingredient of the schemes implemented to cool the internal state of a particle trapped with an atomic ion are (i) the availability of a coupling between the internal and external degrees of freedom of the particle to cool (usually represented as a qubit, i.e., a system with two internal states, \ket{0} and \ket{1}), and (ii) the possibility to control (switch on or off) this coupling. Let us consider a two-particle system, and that on one of the particles (say, particle \#1) we can act with a mechanism to put it into an internal state of our choice (i.e., realize single-qubit operations, for example with a resonant laser) at any moment and that we can also cool its translational motion (e.g., with laser cooling). This allows to internally cool particle \#2 by dissipating its internal energy in the following way (cf.\ Fig.~\ref{fig:cooling}):
\begin{enumerate}
\item At a given moment, switch on the coupling between internal and external d.o.f. of particle \#2, so that the internal excitation energy is transformed into translational energy. Then, switch the coupling off.
\item Particle \#2's translational energy will be shared with particle \#1 due to their interaction, i.e., in the form of excitations of the collective modes of the two-particle system (indicated `$n$=1' in the figure).
\item This excitation energy of the collective modes can then be transferred into internal energy of \#1 switching on and off the coupling mechanism of this particle, in inverse analogy to step 1.
\item Finally, the internal excitation energy of \#1 is taken off the system by reinitializing it to its ground state.
\end{enumerate}
The coupling between internal and external d.o.f. can be realized using state-dependent forces, such as resonant or off-resonant laser pulses: the polarization of the impinging photons can be selected such that only one of the internal states, say \ket{1}, is affected by the laser and hence, can be excited or deexcited.
With an appropriate choice of frequency detuning, $\Delta=\om_{\mathrm{int}}-\om_{\mathrm{L}}$ (with $\om_{\mathrm{int}}=E(\ket{1}_{\#2})-E(\ket{0}_{\#2})$ the internal excitation energy of particle \#2, and $\om_{\mathrm{L}}$ the photon frequency), such as $\Delta=E(n=1)-E(n=0)-\om_{\mathrm{int}}$, transfers the internal excitation energy of \#2 into excitation of the collective mode $n$ (first step in Fig.~\ref{fig:cooling}). As here $\Delta$ is below the ``carrier frequency'', $E(n=1)-E(n=0)$, this is a \textit{red sideband transition}.
Due to the richer internal structure of rotational and vibrational levels in \textit{molecular ions}, the coupling between internal and external d.o.f. for this case is usually more complex. For example, the approach of Ref.~\cite{vogelius2006} was to implement a Raman scheme between an excited $\ket{J=2}$ and the lower $\ket{J=0}$ rotational levels in the ground electronic state of the molecule via an intermediate electronically and rotationally $\ket{J=1}$ excited state.
Ref.~\cite{idzi2011} proposed instead to sympathetically cool a polar molecule with a laser-cooled ion using a time-dependent trapping frequency of the ion to bridge the energy difference between ion and dipole trapping frequencies.

The description developed in Sect.~\ref{sec:eigenmodes} of the ion-dipole system in terms of collective modes allows us to devise a completely analogous procedure to be applied for the internal cooling of \textit{neutral (polar) molecules}, greatly extending the range of systems that can be cooled with atomic (or molecular) ions. Again, we require a mechanism that couples the internal (rotations, vibrations) and external (trapping) d.o.f. of the molecule. We also need that the change in internal state does not lead to untrapping. For molecules confined in dipole traps or optical lattices, the trapping force is determined by the molecule's polarizability, $\alpha$, rather than its dipole moment. Hence, as long as one relies only on internal states with similar $\alpha$, schemes analogous to that used in~\cite{vogelius2006} will be applicable to cool rotational excitations~\footnote{Black-body radiation effects can be incorporated in a straightforward manner with an approach as that in~\cite{vogelius2006}.}.
Further, by using magnetic field gradients or rf radiation to couple different hyperfine states~\cite{mintert2001,johanning09,murpetit2012}, one could also remove internal energy from this degree of freedom as in~\cite{belmechri2013}.


\subsection{Atom-molecule entanglement}
\label{ssec:entanglement}

Collective modes and state-dependent forces have also been used to generate entanglement between two atomic ions and to implement quantum gates between them~\cite{leibfried2003}. In this particular experiment, a Raman-like scheme was implemented between two internal states of $^9$Be$^+$ ions such that if both ions were in the same internal state, both of them were ``pushed'' in the same way by the lasers, while if they were in different states, the stretch mode was excited. Tuning the driving laser's power, it was possible to ensure that, for a given pulse duration, $T$, the accumulated phase was exactly $\pi/2$, thus realizing a controlled $\pi$-phase gate~\cite{leibfried2003}.

\begin{figure}[tb]
  \centering
  \includegraphics[width=0.9\columnwidth]{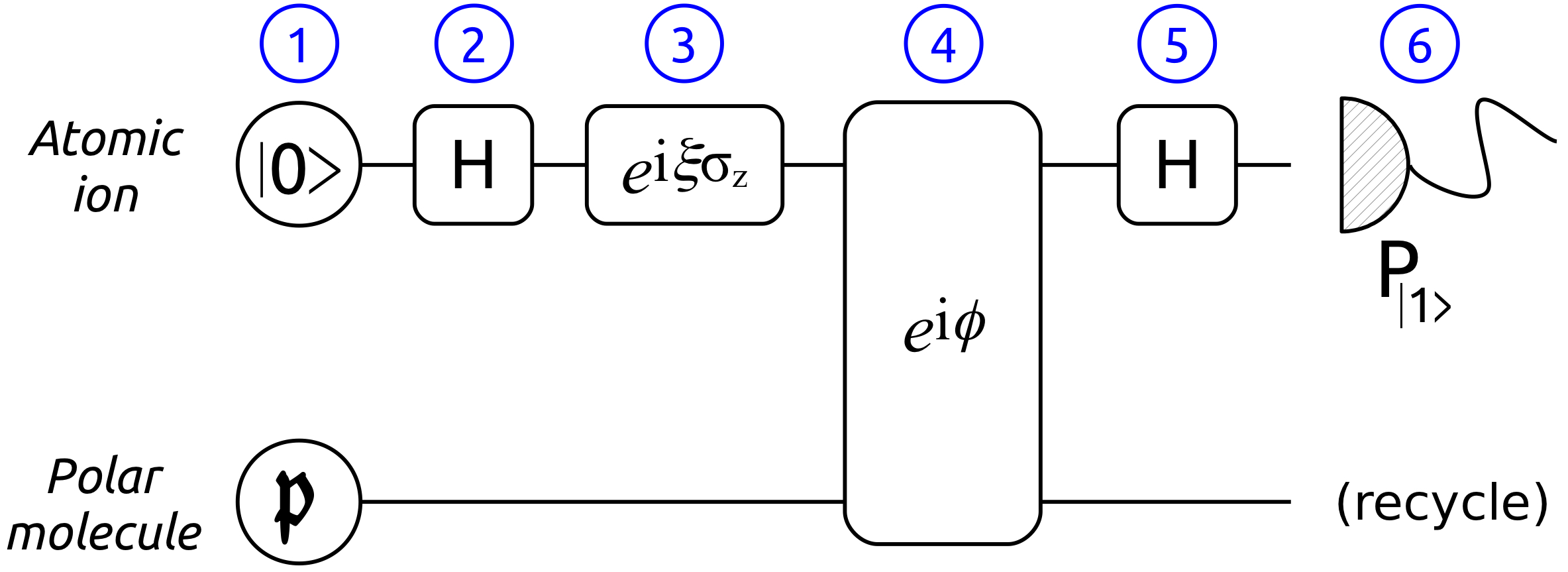}
  \caption{\textit{Entanglement and measurement protocol.}
  The numbers in circles refer to the protocol steps described in Sect.~\ref{ssec:entanglement} (\circledtext{1}-\circledtext{4}) and Sect.~\ref{ssec:edm} (\circledtext{5}-\circledtext{6}).}
  \label{fig:entanglement}
\end{figure}

We have already noted that for polar molecules one can control also, in principle, the coupling of internal and external degrees of freedom.
It is then immediate to design a protocol to entangle a polar molecule with an ion using their collective modes or, conversely, to obtain a measurement of the molecule's EDM by measuring the (rather weak) force between ion and dipole, as we have discussed more deeply in~\cite{iondipoleshort}.
Essentially, the idea follows from applying an interferometric protocol as schematically shown in Fig.~\ref{fig:entanglement}.
The steps here are basically the same as those of the spectroscopy protocol for molecular ions that we have introduced in~\cite{murpetit2012}:
\begin{itemize}
\item[\circledtext{1}] Initialize the atomic ion to its internal ground state, \ket{0}.
\item[\circledtext{2}] Transfer it into an equal superposition of \ket{0} and \ket{1} by means of a $\pi/2$ pulse, i.e., in quantum information parlance, apply a Hadamard gate, $H=(\sigma_x+\sigma_z)/\sqrt{2}$. This amounts to opening the two arms of a Ramsey interferometer~\cite{ramsey1989nobel}.
\item[\circledtext{3}] Discretionally, introduce a reference phase $\xi$ for ion state \ket{1} with respect to \ket{0}.
\item[\circledtext{4}] Apply forces on both ion and molecule, with $f_{\mathrm{ion}}$ being state-dependent. Together with the coupling transmitted by the collective modes, this leads to a state-dependent geometric phase, $\phi$~\cite{leibfried2003,garciaripoll2003,garciaripoll2005,murpetit2012}.
\end{itemize}
At this point (after the two-qubit gate in Fig.~\ref{fig:entanglement}), ion and dipole are entangled.
Indeed, for a separable initial state in the internal space of ion and molecule, 
\begin{subequations}
\begin{eqnarray}
 \ket{in}
 &=& \ket{in}_{\mathrm{ion}}\otimes\ket{in}_{\mathrm{mol}} \:, \\
 \ket{in}_s
 &=&
  \alpha_{\mathrm{s}}\ket{1}_s + \beta_{\mathrm{s}}\ket{0}_s
  = \left( \begin{array}{c} \alpha_{\mathrm{s}} \\ \beta_{\mathrm{s}}   
  \end{array} \right) ,
  \quad s=\mathrm{ion,mol} \,,
\end{eqnarray}
\end{subequations}
after the state-dependent forces are over, the two-par\-ti\-cle state is entangled. To see this, let us consider as an example the case that we implement a controlled phase gate [which reads $C(\phi)=\mathrm{diag}(1,1,1,\exp(i\phi))$ in the basis $\{ \ket{11}, \ket{10}, \ket{01}, \ket{00}\}$, where the first (last) digit indicates the ion (molecule) qubit state]. Then, at the end of step \circledtext{4}, the state reads
\begin{eqnarray}
  \frac{1}{\sqrt{2}} \left( \begin{array}{l}
    (\alpha_{\mathrm{ion}}+\beta_{\mathrm{ion}})
      \alpha_{\mathrm{mol}}e^{i\xi} \\
    (\alpha_{\mathrm{ion}}+\beta_{\mathrm{ion}})
      \beta_{\mathrm{mol}}e^{i\xi} \\
    (\alpha_{\mathrm{ion}}-\beta_{\mathrm{ion}})
      \alpha_{\mathrm{mol}} \\
    (\alpha_{\mathrm{ion}}-\beta_{\mathrm{ion}})
      \beta_{\mathrm{mol}} e^{i\phi} \end{array}\right) \:,
 \label{eq:entangled}
\end{eqnarray}
which is not separable. Apart from other applications discussed below, we mention here that establishing entanglement between ions and polar molecules would constitute a first step towards quantum information processing with these novel kind of hybrid systems.

DeMille was the first to discuss polar molecules as a resource for quantum information processing in a proposal to confine them in an optical lattice embedded inside an inhomogenous electric field allowing to address them individually~\cite{demille2002}.
More recently, \cite{andre2006} considered using their lowest rotational levels as qubit states, and couple them via microwave photons in a hybrid setup with superconducting resonators. On the other hand,~\cite{pellegrini2011} considered encoding the qubit in hyperfine states, and implementing one- and two-qubit operations via microwave pulses.

\jmprevision{
Our proposal here is closer to that in Ref.~\cite{ortner2011}, where a two-dimensional array of polar molecules is proposed for quantum information processing by manipulating each molecule with \textit{another molecule} that is trapped in a parallel layer, all molecules being polarized by a strong perpendicular field. The main difference between the present proposal and~\cite{ortner2011} lies in our relying on the stronger ion-dipole interaction. On the other hand, a setup with only polar species and no ions might be easier to implement, as one does not need to consider the cross-talk between ion and molecule traps. For the case of the ion-dipole setup, one would in addition benefit from the high degree of control on internal and external degrees of freedom already demonstrated in experiments.
}


\subsection{Measuring and controlling EDMs}
\label{ssec:edm}

The internal-state entanglement achieved with the previous protocol is encoded in the phase $\phi$ of the two-qubit gate. This phase depends on the fields applied to the two particles (strength, duration) \textit{and} on the frequencies of the normal modes. Hence, if we apply known fields to the two particles, a measurement of the phase accumulated, $\phi$, provides information on these modes or, in other words, on the ion-dipole coupling, $\alpha$.
We can obtain the value of the phase by transforming the information encoded in the phase differences between the basis states in Eq.~(\ref{eq:entangled}) into population information. To this end, after the two-particle gate, one applies a new $\pi/2$ pulse on the ion (which closes the two arms of the Ramsey interferometer) and measures its final state, corresponding to steps \circledtext{5} and \circledtext{6} in Fig.~\ref{fig:entanglement}.

When the forces applied on the ion and the molecule are given by
\begin{eqnarray}
 f_{\mathrm{s}}(t)=\fo{s}e^{-(2t/T)^2} \cos(\om_{\mathrm{dr}}t) ,
 \quad \mathrm{s=ion,mol}
 \:,
 \label{eq:driving}
\end{eqnarray}
the probability that the ion is found in \ket{1} depends on the phase $\phi$  through (cf.~\cite{murpetit2012})
\begin{subequations}
\begin{eqnarray}
 \phi
 &=&
 \sum_{n=\mathrm{com,str}} \beta_n \phi_n \:, \\
 \phi_n
 &=&
 \sqrt{\frac{\pi}{2}}\frac{\fo{ion}\fo{mol} a_{n}^2 T}
 {4\om}\Xi \:,
 \\
 \Xi
 &:=&
 \frac{\om^2}{\omcom^2-\om_{\mathrm{dr}}^2} 
 - \frac{\om^2}{\omstr^2-\om_{\mathrm{dr}}^2} \:,
\end{eqnarray}
\end{subequations}
where $a_n$ is the harmonic oscillator length of the excited collective mode
and $\beta_n(\mu,\nu,\talpha)$ are dimensionless constants.
Now, knowledge of the fields applied on both particles, together with the precise knowledge available on the structure of trapped ions provide \fo{ion}, leaving $\dip$ --which appears in $\omcom,\omstr$ and, at least parametrically, in $\fo{mol}$-- as the only free parameter to be determined. Measurements of $\phi$ can thus be used to obtain \dip\ (see~\cite{iondipoleshort} for more details).
Adding to this, quantum chemistry calculations of molecular structure can provide the missing link between the applied fields, $\dip$ and the molecular levels, closing the loop between the measured phase $\phi$ and other molecular properties~\cite{murpetit2012,murpetit2013}.

In a complementary way, and similarly to what has already been realized for atomic ions~\cite{leibfried2003}, if the EDM of the molecule and the ion-dipole distance are known with enough precision, the same protocol can be employed to realize controlled operations on its internal state and, in particular, its EDM. In this case, one would fix the parameters of the force applied on the molecule, $(\fo{mol},T,\om_{\mathrm{dr}})$, so as to obtain the desired phase on the two-particle state.


\subsection{Measurement uncertainty estimates}
\label{ssec:uncertainty}

To get a feeling of the dependence of the phase $\phi$ on $\dip$ in practical cases, let us first consider a small value of the coupling, $\talpha=10^{-3}\ll 1$. Then, for the case $\mu=\nu=1$, from Eqs.~(\ref{eq:eigenfreqs}) we obtain $\omcom=\om$ (independent of \dip) while $\omstr\approx\om\sqrt{1+60\talpha}$: the dynamics is determined by the \str\ mode.
Hence, using $\phi\propto a_{\mathrm{str}}^2 \propto 1/\omstr$, we arrive at
\begin{eqnarray}
 \frac{d\phi}{d\dip}
 &=&
 \frac{d\talpha}{d\dip}\frac{d\phi}{d\talpha}
 \approx \frac{d\talpha}{d\dip} \frac{-30}{1+60\talpha}\phi
 = \frac{-30\talpha}{1+60\talpha}\frac{\phi}{\dip}
 \nonumber
 \\
 &\Rightarrow &
  \left|\frac{\delta\dip}{\dip}\right|
  \approx
  \left|\frac{1}{30\talpha}\frac{\delta\phi}{\phi}\right|
  \:.
\end{eqnarray}
This means that if we can measure $\phi$ with a relative uncertainty of $\epsilon_\phi$, then we can detect relatives changes in \dip\ of order $\epsilon_{\dip}\approx\epsilon_\phi/(30\talpha)\gg\epsilon_\phi$. Interestingly, the sensitivity increases for smaller $\talpha$ (corresponding to a smaller \dip\ for the same distance $z_0$).

On the other hand, for a case with $\talpha\gtrsim0.1$, when the limiting behaviours corresponding to $\talpha\to\infty$ apply (cf.\ Fig.~\ref{fig:modes}b), the dynamics is dominated by \com, and we have $\phi\propto 1/\omcom\propto1/\sqrt{\talpha} \propto 1/\sqrt{\dip}$. Then,
\begin{equation}
 \frac{d\phi}{d\dip}
 = \frac{-\phi}{2\dip} \:,
\end{equation}
which indicates that the uncertainty we have estimating \dip\ is of the same order that we have measuring $\phi$.

\jmprevision{
The main sources of experimental uncertainty might appear to be related to temperature and alignment.
First, as long as the harmonic approximation for the displacements remains valid, the geometric character of the phase $\phi$ ensures that temperature should not be a concern~\cite{murpetit2012,leibfried2003,garciaripoll2003,garciaripoll2005}.
With respect to the alignment of \dipvec\ with the $z$ axis defined by the ion position, we see from Fig.~\ref{fig:coupling} (top inset) that the effective potential is very smooth around the minimum, and one needs a displacement along $r_{\perp}$ of magnitude $\gtrsim z_0/10$ to be sensitive to this source of error. When this happens, the exact decoupling between $(x,y,z)$ will no longer hold. This can be seen as an effective coupling between the \com\ and \str\ modes in $(x,y)$ with those in $z$. Such effects should only become apparent for times $\gtrsim \om^{-1}\sqrt{z_0/(r_{\perp}\talpha)} \gg \om^{-1}$ for $r_{\perp}\ll z_0$.
}


\subsection{Mapping polar surfaces and ordered phases of dipole lattices: Ion-Dipole Force Microscopy (IDFM)}
\label{ssec:idfm}

\begin{figure}[bt]
  \centering
  \includegraphics[width=0.9\columnwidth]{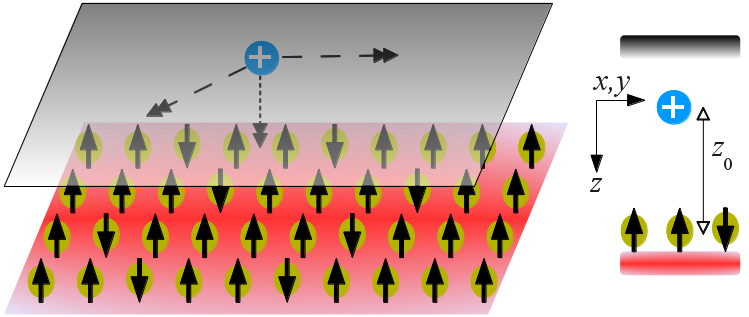}
  \caption{{\em Ion scanning a lattice of dipoles.}
  (Left) An ion (blue circle) is trapped close to a microfrabricated surface trap (grey shaded area) which allows to move it in $x,y$ directions (dashed arrows) above a lattice of dipoles (thick arrows in yellow ovals), and also to shift it closer to the dipoles (dotted arrow).
  (Right) Cross section showing coordinate system.}
  \label{fig:lattice}
\end{figure}
As a last potential application, we mention the mapping of EDMs arranged on two-dimensional lattices and, complementarily, the creation of ``topographical'' maps of surfaces covered with polar molecules.
\jmprevision{%
The experimental setup in the first case would consist of a two-dimen\-sion\-al arrangement of electric dipoles, such as polar molecules confined in an optical lattice ($x$-$y$ plane), and a trapped ion hovering over them, which can be displaced along the $x$, $y$, and $z$ directions, see Fig.~\ref{fig:lattice}.
One could distribute polar molecules in such regular arrays by trapping them in sufficiently deep optical lattices;
experiments with KRb molecules trapped in 2D and 3D optical lattices have already been reported from several laboratories, see e.g.~\cite{ospelkaus2006,demiranda2011}.
In these conditions, theoretical calculations have shown that the dipole-dipole interaction leads to exotic quantum phases~\cite{gorshkov2011,goral2002,barnett2006}, whose direct observation is however lacking. Here, we propose to use an atomic ion to probe such molecular systems.
}

In an experimental sequence, one would first transport the ion along $x,y$ as in ion-shuffling experiments~\cite{hensinger2006,blakestad2009,maiwald2009,%
moehring2011,wright2013} at a constant, large value of $z \gg L$, to the desired position above the lattice, and then shuttle the ion down to $z \sim L \sim 1~\mumeter$ for a strong modification of the coupling strength, $\talpha\propto 1/z^4$, so that in practice only the nearest EDM situated below the ion is relevant to determine its response to external drivings~(\ref{eq:driving}). This scheme would enable a sensitive measurement of EDM values in an approach similar to atomic force microscopy, but at larger distances from the surface under study. This may be a useful approach to investigate the formation of ordered magnetic phases in ultracold samples~\cite{gorshkov2011,lahaye2009} without recurring to light polarization-analysis techniques based on the magnetico-optic Kerr effect (cf.~\cite{qiu2000,lehnert2009} for reviews from a surface science veiwpoint) similar to the proposal in Ref.~\cite{eckert2007} to study spinor gases.

\begin{figure}[bt]
  \centering
  \includegraphics[width=\columnwidth]{
     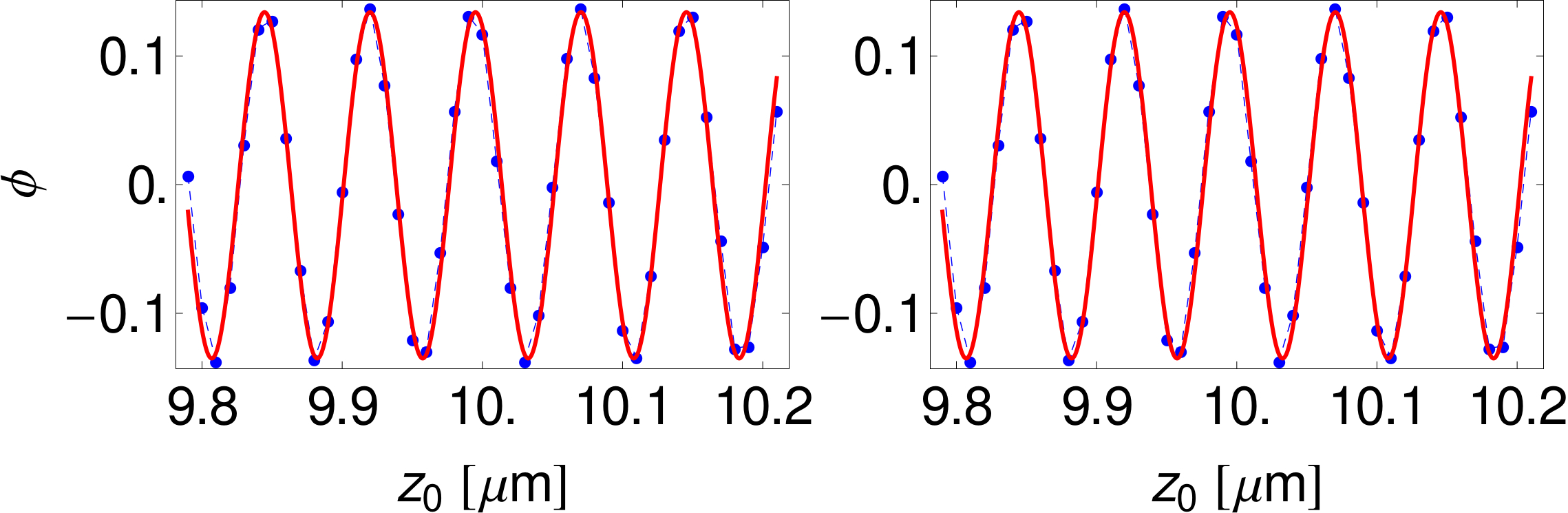}
  \caption{%
  \textit{Ion-dipole force microscopy.}
  Accumulated phase vs.\ ion-dipole distance
  $z_0\approx20~\mumeter$ (left) and $z_0\approx10~\mumeter$ (right)
  for a Ca$^+$-KRb system.
  Blue circles are calculated values, and the red line is a sinusoidal fit through the data.
}
  \label{fig:idfm}
\end{figure}

To get a rough estimate of the spatial resolution that one might achieve with this procedure, we show in Fig.~\ref{fig:idfm} the phase accumulated by applying the interferometric measurement protocol in Fig.~\ref{fig:entanglement}, for the case of a $^{40}$Ca$^+$ ion on a $^{40}$K$^{87}$Rb molecule in its ground rovibronic state, $\ket{\mathrm{X}{}^1\Sigma,v=0,J=0}$, which has an EDM $\dip_\mathrm{KRb}=0.566$~D~\cite{ni2008sci}, when they are harmonically confined with trapping frequencies as in Table~\ref{tab:frequencies} (corresponding to $L=4.3~\mumeter$), and separated a distance $z\approx 20~\mumeter$ (left; $\alpha/\hbar\sqrt{\om\Om}=0.3$) and $10~\mumeter$ (right; $\alpha/\hbar\sqrt{\om\Om}=1.3$).
\jmprevision{
Under these conditions, the expected molecule displacement caused by approaching the ion is $(z_*-z_0)\sim 10^{-3}z_0 \sim 20$~nm, see Fig.~\ref{fig:displacement}.
Hence, the
}
smooth dependence on $z_0$ observed in Fig.~\ref{fig:idfm} for both cases would allow sub-$\mumeter$ resolution measurements.
From our simulations, the main limiting factors of this technique appear to be related to excited-state decay of ion and molecule and photon scattering from the off-resonant pulses, which can in principle be greatly reduced using two-photon Raman schemes~\cite{vogelius2006}
(see also~\cite{murpetit2012,cadarso2013arxiv}) to minimize the population in the excited state.
\jmprevision{
As before, temperatures for which the thermal displacements are small should be no concern, as well as the ion micromotion.
}

The previous application of IDFM requires a previous knowledge of the ion-surface distance $z_0$. How to determine this variable is therefore necessary. Actually, IDFM can be used to determine this height, which would enable the creation of relief maps of polar surfaces. To this end, a useful protocol would rely on the measurement of the collective mode frequencies $\omcom$,$\omstr$, e.g., via a resonant-excitation mechanism as used for molecular mass spectrometry in~\cite{drewsen2004}
(other approaches have also been realized, see e.g.,~\cite{goeders2013,iondipoleshort,haeffner2008}).
Given that the ratio $\omstr/\omcom$ is a known function of the parameters
($\mu,\nu,\talpha$)  [Eqs.~(\ref{eq:eigenfreqs})], it is then possible to determine $\talpha$ and hence $z_0$, if $\dip$ is known.
If $\dip$ is unknown, this protocol would enable to measure surface relief up to a rescaling factor, which could be determined independently.
Finally, the same protocol can be used to measure \textit{both $\dip$ and $z_0$} by performing it at a set of distances: as the coupling $\talpha$ depends linearly on \dip\ but quadratically on $z_0$, it would be possible to determine both unknowns from a multivariate fit to the data.

\jmprevision{
In the abscence of a periodic potential, the dipole-dipole interaction can lead to a quantum phase transition into a crystalline phase~\cite{buechler2007,astrak2007}. We do not discuss this situation here, as it would require an independent analysis of the excitation modes of the crystal due to the presence of the ion and its interaction with the molecules. In addition, usually the self-assembly relies on the presence of a (strong) electric field to orient the dipoles, which would render the trapping and control of the nearby ion rather difficult.
}


\section{Discussion and Outlook}
\label{sec:outlook}

In summary, we have studied a simple model of an atomic ion and a polar molecule trapped in a single hybrid setup, taking into consideration their electrostatic interaction.
We have determined the displacement of their initial equilibrium positions due to this interaction, and the resulting normal modes of collective excitation, which we have obtained in analytical form.
We have discussed a few possible applications that build on the knowledge of these collective modes, from internal-state cooling of molecules, to establishing entanglement between molecules and ions and realizing two-qubit gates between them, to studying the distribution of dipole moments on two-dimensional arrays of polar molecules in optical lattices, a setup proposed to quantum-simulate various strongly-correlated models of condensed matter.

\jmprevision{
As limitations for protocols discussed, we have discussed its resilience to (small) thermal fluctuations. We have also given an estimate of the timescale upon which a misalignment between the ion-molecule axis and the dipole might trigger effects due to couplings that require a numerical treatment beyond the analytical description presented here. 
The main challenge remains to build a hybrid setup to trap an ion and a molecule, so that each particle's trap does not adversely perturb the other, a subject on which we have provided some realistic proposals.
}

Our work is framed in ongoing efforts to build hybrid quantum systems with diverse quantum technologies\textemdash atoms, ions, quantum circuits, cold molecules, photonic fibers, etc~\cite{xiang2013}. These systems are expected to combine the best properties of each constituent element while avoiding some of their main drawbacks. In this context, cold molecules offer a rich internal structure with a large number of levels covering a broad range of energies, from kHz to GHz and beyond. Thus, they appear as attractive candidates for frequency conversion between otherwise ``incompatible'' components. Long-lived rotational levels of ground-state homonuclear molecules such as O$_2^+$ have also been identified as good candidates for storage of quantum information due to their almost perfect shielding from environmental perturbations such as black-body radiation or magnetic field fluctuations~\cite{murpetit2013} (see also \cite{yun2013}).

Many other applications for cold molecules, especially polar molecules, have been proposed for both fundamental and applied studies~\cite{carr2009}, and we expect that our contribution here will help to establish links of this developing technology with more advanced ones such as trapped atomic ions.
In conclusion, we would like to think of the present work as a foundational stone for future joint applications of trapped ions and molecules, following the milestone work by Wolfgang Paul on the trapping and control of particles at the quantum, single-particle level.

\begin{acknowledgements}
This work supported by Spain MINECO Proj\-ect FIS2012-33022 (Spain), 
CAM research con\-sor\-tium QUITEMAD (S2009-ESP-1594),
ESF COST Action IOTA (MP1001),
EU FP7 networks PROMISCE and POLATOM,
US NSF 
(Grant No.\ NSF PHY11-25915),
an EU FP7 Marie Curie fellowship (MOLOPTLAT), and the CSIC JAE-Doc Program co-funded by European Social Fund (EU).
\end{acknowledgements}

\bibliographystyle{spphys} 

\end{document}